# Ozone Abundance in a Nitrogen-Carbon Dioxide Dominated Terrestrial Paleoatmosphere

Brian C. Thomas[1], Adrian L. Melott[1], Larry D. Martin[2], and Charles H. Jackman[3]

Correspondence:
Adrian L. Melott
Department of Physics and Astronomy
University of Kansas
1251 Wescoe Dr. # 1082
Lawrence, KS 66045-7582

Phone: 785 864 3037
Fax: 785 864 5262
Email: melott@ku.edu

1. Department of Physics and Astronomy, University of Kansas, Lawrence, KS
2. Biodiversity Center and Department of Ecology and Evolutionary Biology, University of Kansas, Lawrence KS
3. Laboratory for Atmospheres, NASA Goddard Space Flight Center, Code 916, Greenbelt, MD

# ABSTRACT

We compute the ozone distribution for a model terrestrial paleoatmosphere in which the present oxygen abundance is largely replaced by carbon dioxide, which we argue is a reasonable working assumption. In principle, the presence of carbon dioxide might supplement the ozone shield as compared with models based on nitrogen without high carbon dioxide abundance so that early life need not have been as UV-resistant as often assumed. An extrasolar planet with a high-$CO_2$ atmosphere might contain enough $O_3$ to be a source of false positive biomarkers. We find that the globally averaged $O_3$ column density can be the same, or nearly four times higher (depending upon the $O_2$ partial pressure) when $CO_2$ is used in place of $N_2$ as the replacement component for lowered $O_2$ in a 1-atm terrestrial planet with solar radiation. The effect is important for making quantitative deductions from future data, but does not invalidate the use of $O_3$ as a biomarker for free oxygen. These results make prospects for detection of extrasolar planetary $O_3$ absorption somewhat better than before.

INTRODUCTION

Ozone ($O_3$) is a proposed biomarker gas, because it is produced as a by-product of molecular oxygen ($O_2$), a presumed product of photosynthesis for life with chemistry similar to that which dominates our biosphere. Proposed space missions (Beichman et al. 1999; Léger 2000) are to take spectra of the atmospheres of extrasolar planets. For this reason, the abundance of ozone in plausible nonbiogenic atmospheres is important.

Also, ozone acts as a shield against ultraviolet radiation, particularly the UVB band (280-315 *nm*), where DNA absorbs strongly and is damaged. The destruction of the ozone shield by a variety of astrophysical events (e.g. Cockell & Blaustein 2000; Melott et al. 2004 and references therein) has been presented as a plausible mechanism for certain mass extinctions that have impacted the Earth. Supernovae and gamma-ray burst rates are thought to be coupled to the star formation rate, and therefore to have been more common in the past. For this reason, the extent to which Paleoproterzoic life was shielded against UV, and presumably therefore likely to have been sensitive to it, is important in assessing the likely biological impact of any episodic ozone depletion at that time.

Recently, Segura et al (2004) examined ozone concentrations and the resulting UV fluxes on Earthlike planets around near-Solar stars. Concentrations of a variety of constituents were reported for several stellar spectral types and a range over several orders of magnitude of oxygen concentration (relative to PAL). Most importantly for our work here, when the atmospheric composition was varied, $O_2$ was "replaced" by $N_2$. This may potentially affect the results. Even low levels of these molecules are potential biomarkers, as they are likely characteristic of a terrestrial exoplanet atmosphere after the onset of photosynthesis, but before "oxidation" of the planet (Kasting & Catling 2003).

The early atmosphere is not well understood. However, we wish to argue the value of considering the same issues for a high $CO_2$ atmosphere. The photolysis of $CO_2$ may possibly contribute to the ozone abundance, since molecular oxygen is a probable product. There are a variety of motivations for considering the high-$CO_2$ case. First, the nearly terrestrial planets of Venus and Mars have atmospheres now dominated by $CO_2$. Secondly, recent evidence from siderite beds (Ohmoto et al. 2004) and carbon isotope ratios in microfossils (Kaufman and Xiao 2003) indicate a $CO_2$-rich atmosphere before 1.8 Gya. Thirdly, the Earth contains about $10^{20}$ kg of carbon, mostly in the form of sediments (Schlesinger 1997, and references therein). Fourthly, photosynthesis results in a net conversion of $CO_2$ to $O_2$ with some carbon burial.

$CO_2$ is continually outgassed from the crust, and presumably was similarly contributed to the atmospheres of the other terrestrial planets. Apparently photosynthesis may have begun quite early (Kasting & Catling 2003; Tice & Lowe 2004), but free $O_2$ levels in the atmosphere remained quite low until the ready supply of inorganic reactants were oxidized. The net rate of carbon burial was rather low until about the time substantial $O_2$ began to appear (Schlesinger 1997). We therefore propose to examine an alternative model atmosphere: lowered levels of $O_2$, at 1 atm total pressure, but with the $O_2$ "replaced" by $CO_2$, keeping the partial pressure of $N_2$ fixed near its present level. We do not pretend that this is an exhaustive exploration of parameter space. Our goal is a preliminary assessment of the significance of $CO_2$ as a possible ozone progenitor. For this reason, we restrict ourselves to the Solar spectral type for now. We note that $CO_2$ is not a significant UV absorber (Liou 2002), so that $O_3$ is still required for significant shielding in this case.

ATMOSPHERIC COMPUTATIONAL MODEL

We use the Goddard Space Flight Center two-dimensional code originally described in Douglass, Jackman, & Stolarski (1989), with further modifications and improvements as described in Jackman et al. (1990). The two dimensions are latitude (-90° to +90°) and altitude up to 116 km. The latitude is divided in to 18 bands of 10 degrees. The altitude range includes 58 levels equally spaced in log pressure (appx. 2 km each). Heterogeneous processes in the stratospheric sulfate aerosol layer and polar stratospheric clouds are treated as in Considine, Douglass, & Jackman (1994). A lookup table is used for computation of the photolytic source term in calculations of photodissociation rates of atmospheric constituents by sunlight (Jackman et al. 1996). Winds and small-scale mixing are included as described in Fleming et al. (1999). Our time-dependent code follows changes through day/night cycles as well as annual effects due to the obliquity of the Earth.

As compared with the uses of this code cited above, anthropogenic components such as CFCs have been removed for the computation reported here. Since we restrict our attention to a first-order computation of the effect of a $CO_2$-rich atmosphere on the ozone signature as compared with an $N_2$-rich atmosphere, we have removed the 11 year solar cycle from the code. As a rough analogy to comparison with the results of Segura et al.

(2003), which are one-dimensional (do not include latitude, and by implication latitudinal mixing), and are time-independent at a fixed solar angle of 45°, we report global averages over an entire year. We note that our semi-empirical code is based on measured values of temperature, wind currents, etc. and can only be considered as a qualitative guide to the expected behavior of terrestrial planets. Segura et al. compute a self-consistent altitudinal temperature profile. Finally, we do not study detailed spectra of such planetary atmospheres, but can report on whether the presence of $CO_2$ makes a substantial difference in the ozone content as compared with $N_2$.

RESULTS

First, our results confirm those of Kasting et al.(1984); see also Kasting & Catling (2003), as against those of Selsis et al. (2002): photolysis of a high-$CO_2$ atmosphere does not generate large amounts of $O_3$, sufficient to mimic the effects of considerable photosynthesis. However, we do see differences between the $CO_2$ and $N_2$ replacement cases. Our results are summarized in Fig. 1, where we show global average $O_3$ column density as a function of $O_2$ abundance.

For $O_2$ at 1 PAL (present atmospheric level), our result shown in Fig. 1 ($N_2$) agrees with Segura et al. (2004) to better than 4%. Given the considerable difference in computational approach, this agreement is probably fortuitous. Our results vary from theirs for lower $O_2$ partial pressure, but the disagreement (factor of a few) is not surprising given the major differences in the assumptions used in the computations.

More significant is the relative difference in the $O_3$ column density between our nitrogen-oxygen atmosphere and our nitrogen-carbon dioxide-oxygen atmosphere. We did the computation for the $O_2$ values shown in Fig. 1 as well as for $10^{-13}$, although we did not plot the latter in the interest of legibility. The plot shows values possible for a photosynthetic biotic atmosphere, and $10^{-13}$ is typical of what is expected in a prebiotic atmosphere (Kasting and Catling 2003). Over a broad range from $10^{-13}$ to $10^{-1}$ PAL $O_2$, we find that the $O_3$ column density is greater in the presence of carbon dioxide, up to nearly a factor of 4 larger than with nitrogen. For $O_2$ at $10^{-13}$ PAL, we have $9 \times 10^{11}$ cm$^{-2}$ and $4 \times 10^{11}$ cm$^{-2}$ for the $O_3$ column density in the carbon dioxide and nitrogen cases respectively.

DISCUSSION

We have argued that nitrogen-carbon dioxide atmospheres with free oxygen contamination are a likely model for terrestrial type planetary atmospheres before the oxidation of the biosphere than in the absence of carbon dioxide. We have done an assessment of the globally averaged ozone column density in 1-atm pressure, solar irradiated atmospheres. We find that our model atmospheres with high $CO_2$ content produce up to about 4 times the globally averaged $O_3$ column density. This makes a substantive difference to the result, but does not invalidate the use of $O_3$ as a biomarker for free oxygen. The levels of $O_3$ produced in our model make a quantitative, but not qualitative difference to the onset time of UVB shielding as photosynthesis began to strongly modify the atmosphere. Any surface life on Earth prior to the rise of $O_2$ can be assumed to have been relatively UV-hardened.

$O_3$ continues to appear to be a valid marker for photosynthesis driven, water-carbon based life on terrestrial planets. Our results suggest that inclusion of $CO_2$ will make this conclusion stronger, and the prospects of detection somewhat better.

ACKNOWLEDGMENTS

We gratefully acknowledge the support of NASA Astrobiology grant NNG04GM41G, and the advice of Rich Stolarski on subtleties of dealing with $CO_2$.

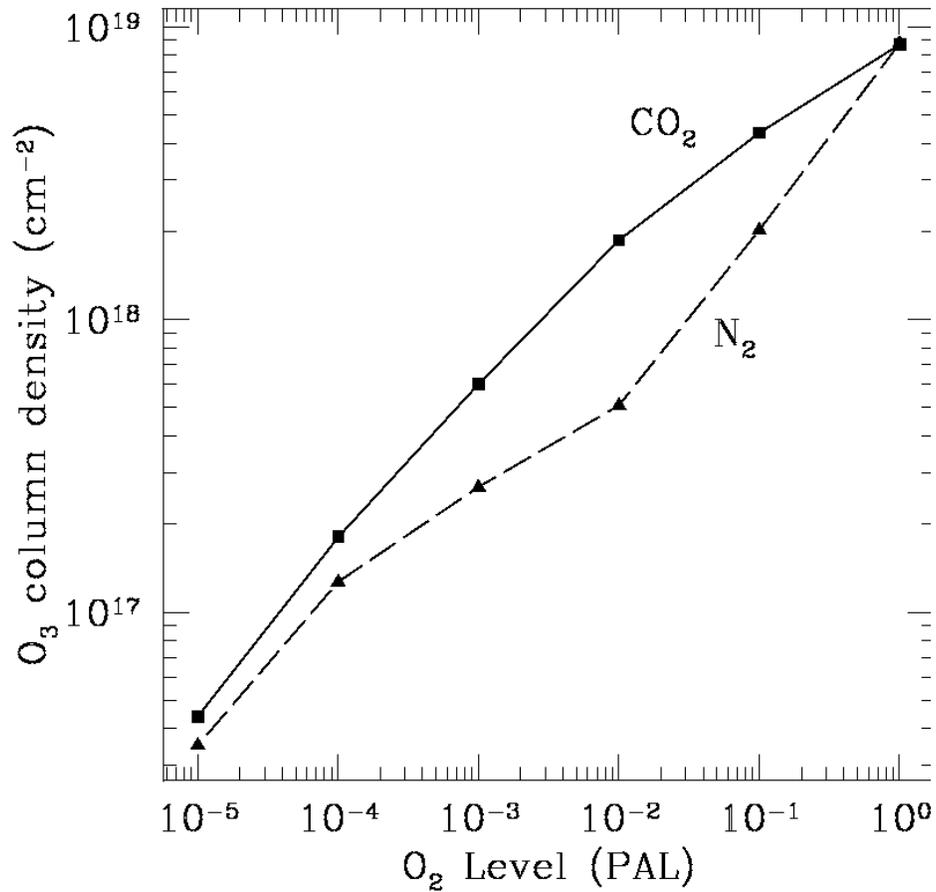

FIGURE CAPTION

We show the annually and globally averaged column density of $O_3$ as a function of the $O_2$ abundance (in units of the Present Atmospheric Level) for two model atmospheres. Both contain a minimum of the present PAL of $N_2$ and $CO_2$. The solid line shows the result if the missing $O_2$ is replaced by $CO_2$; the dashed line shows the result if it is replaced by $N_2$ instead.